\documentclass[12pt]{article}

\usepackage{amsmath}
\usepackage{amssymb}
\usepackage{amsfonts}
\usepackage[mathscr]{euscript}

\title{\bf Extended Reduction Criterion and Lattice States}

\author{Fabio Benatti$^{a,b}$, Roberto Floreanini$^{b}$,
Alexandra M. Liguori\\
\small $^a$Dipartimento di Fisica Teorica, Universit\`a di Trieste,
Strada Costiera 11,\\
\small 34014 Trieste, Italy\\
\small ${}^b$Istituto Nazionale di Fisica Nucleare, Sezione di Trieste,
34100 Trieste, Italy}

\date{\null}

\begin{document}

\maketitle

\vskip 2cm

\begin{abstract}
\noindent We study a particular class of states of a bipartite
system consisting of two $4$-level parties. By means of an adapted
extended reduction criterion we associate their entanglement
properties to the geometric patterns of a planar lattice consisting
of $16$ points.
\end{abstract}

\section{Introduction}
\label{intro}

In recent years, due to the rapid growth of quantum information,
communication and computation, the necessity of identifying,
quantifying and classifying entangled states as a physical resource
has given birth to Quantum Entanglement Theory (see the
review~\cite{horo2004} and references therein). Though the focus is
now shifting to multipartite entanglement, still generic quantum
correlations in finite dimensional bipartite systems are not
completely understood.

From a mathematical point of view, the lack of exhaustive criteria
capable of witnessing the presence of bipartite entanglement, is
nothing else but the absence of a complete characterization of
positive, but not completely positive maps~\cite{C,W,S}. Looking for
new classes of entangled states~\cite{koss1,koss2} and for dedicated
entanglement witnesses~\cite{alb} may thus improve the comprehension
of the algebraic structure behind physical phenomena like
separability, free and bound entanglement~\cite{horo2004}.

Recently, an extension of the so-called reduction map~\cite{horo99}
has been proposed~\cite{breuer1} that is tailor-made for revealing
the entanglement of states with a specific structure with respect to
angular momentum. Subsequently, this map was shown~\cite{hall} to be
a particular instance of a larger class of indecomposable positive
maps~\cite{C,W,S}. In the following, we shall adapt the extended
reduction criterion to the study of the class of bipartite states of
two $4$-level parties which were introduced in~\cite{ben-flore3} and
further studied in~\cite{ben-flore2}. For reasons that will soon
become obvious, these $16\times 16$ density matrices have been
called lattice states.

These states are characterized by subsets of a discrete planar
lattice with $16$ elements and can be grouped together into
equivalence classes characterized by specific geometric patterns. By
means of the extended reduction map, we shall associate most of them
with specific properties of the corresponding lattice states like
that of being separable, NPT-entangled or PPT-entangled. Because of
this, besides enriching the phenomenology of entangled states, the
class of lattice states may also turn out to be a useful arena for
the approach based on the discrete Wigner
functions~\cite{GHW,PR1,PR2,Galv}

The plan of the paper is as follows: we shall first introduce
the lattice states and their presently known properties; then,
we shall improve their classification by using the extended
reduction criterion; finally, we shall completely characterize
some subclasses of them and discuss which ones need stronger
entanglement witnesses than those available so far.

\section{Lattice states}
\label{lattice_states}

The construction of lattice states is as follows.
Let $\sigma_\alpha$, $0\leq\alpha\leq3$,
be the Pauli matrices supplemented with $\sigma_0=I_2$, the $2\times 2$
identity matrix, and consider the totally symmetric state
$$
|\Psi_+^4\rangle=\frac{1}{2}\sum_{i=1}^4|ii\rangle\in \mathbf{C}^{16}\ ,
$$
where $\{|i\rangle\}_{i=1}^4$ is a fixed orthonormal basis in
$\mathbf{C}^4$.
Denoting the tensor products of pairs of Pauli matrices by
\begin{equation}
\label{Pauli-tensprod}
\sigma_{\alpha \beta} := \sigma_\alpha \otimes \sigma_\beta\ ,
\end{equation}
the action of $I_4\otimes\sigma_{\alpha\beta}$, $I_4=I_2\otimes I_2$,
on $|\Psi^4_+\rangle$ yields an orthonormal basis
$$
|\Psi_{\alpha \beta} \rangle := (I_4 \otimes \sigma_{\alpha \beta})
|\Psi_+^4 \rangle \in\mathbf{C}^{16}\ , $$
with corresponding orthogonal projectors
\begin{equation}
\label{latt-proj}
P_{\alpha \beta} := |\Psi_{\alpha \beta} \rangle \langle
\Psi_{\alpha \beta}| =
(I_4 \otimes \sigma_{\alpha \beta})\,|\Psi_+^4\rangle\langle\Psi_+^4|\,
(I_4 \otimes \sigma_{\alpha \beta})  \ .
\end{equation}
Let $ L_{16}$ denote the finite square lattice consisting of $16$
points labeled by pairs $(\alpha,\beta)$:
$$
L_{16}:=\Bigl\{(\alpha,\beta) : \alpha,\beta = 0,1,2,3 \Bigr\}\ .
$$
We shall also denote by $\mathcal{C}_\alpha$ and $\mathcal{R}_\beta$
the columns and rows of the lattice $L_{16}$:
$$
\mathcal{C}_\alpha=
\Bigl\{(\alpha,\beta)\in L_{16}\,:\, \beta=0,1,2,3\Bigr\}\ ,\quad
\mathcal{R}_\beta=\Bigl\{(\alpha,\beta)\in L_{16}\,:\, \alpha=0,1,2,3\Bigr\}\ .
$$

To every subset $I\subseteq L_{16}$ of cardinality $N_I$, we
associate a mixed state equidistributed over the orthogonal
projectors $P_{\alpha\beta}$ labeled by pairs in $I$:
$$
\varrho_I = \frac{1}{N_I} \sum_{(\alpha,\beta)\in I} P_{\alpha \beta} \ .
$$
We shall refer to the $\varrho_I$ as lattice states: they
are a particular subclass of the mixed states
\begin{equation}
\label{diag-states}
\varrho_\pi=\sum_{(\alpha,\beta)\in L_{16}}\, \pi_{\alpha\beta}\,
P_{\alpha\beta}
\ ,\quad\pi_{\alpha\beta}\geq 0\ ,\ \sum_{(\alpha,\beta)\in
L_{16}}\pi_{\alpha\beta}=1
\end{equation}
that commute with the projectors $P_{\alpha\beta}$.

It turns out that properties of these states like that of being
PPT or bound entangled are related to certain specific geometrical
patterns of the subsets that label them.
Indeed, the following results have been proved
(see~\cite{ben-flore2} and~\cite{ben-flore3}):
\bigskip

\noindent
\textbf{Proposition 1}
\smallskip

\begin{enumerate}
\item
A nececessary and sufficient condition for a lattice state $\varrho_I$ to be
PPT is that for every $(\alpha,\beta)\in L_{16}$
the number of points on $\mathcal{C}_\alpha$ and $\mathcal{R}_\beta$
different from $(\alpha,\beta)$ and beloging to $I$ be not greater than
$N_I/2$. In terms of the characteristic functions
$\chi_I(\alpha,\beta)=1$ if $(\alpha,\beta)\in I$, $=0$ otherwise, a lattice
state $\varrho_I$ is PPT if and only if for all $(\alpha,\beta)\in L_{16}$
\begin{equation}
\label{prop1a}
\sum_{\delta=0,\delta\neq\beta}^3\chi_I(\alpha,\delta)\,+\,
\sum_{\gamma=0,\gamma\neq\alpha}^3\chi_I(\gamma,\beta)\leq\frac{N_I}{2}\ .
\end{equation}
\item
A sufficient condition for a PPT lattice state $\varrho_I$ to be
entangled is that there exists at least a pair
$(\alpha,\beta)\in\ L_{16}$ such that only one point on
$\mathcal{C}_\alpha$ and $\mathcal{R}_\beta$ and different from
$(\alpha,\beta)$ belongs to $I$. Equivalently, $\varrho_I$ is entangled if
a pair $(\alpha,\beta)\in L_{16}$ exists such that
\begin{equation}
\label{prop1b}
\sum_{\delta=0,\delta\neq\beta}^3\chi_I(\alpha,\delta)\,+\,
\sum_{\gamma=0,\gamma\neq\alpha}^3\chi_I(\gamma,\beta)=1\ ,\quad
\hbox{and}\quad (\alpha,\beta)\notin I\ .
\end{equation}
\end{enumerate}
\bigskip

\noindent
\textbf{Remarks 1}\hfill

\begin{enumerate}
\item
A necessary and sufficient condition for PPTness can be worked out
for the more general states in~(\ref{diag-states}); it reads:
\begin{equation}
\label{diag-states-PPT}
\sum_{\delta=0,\delta\neq\beta}^3\pi_{\alpha\delta}\,+\,
\sum_{\gamma=0,\gamma\neq\alpha}^3\pi_{\gamma\beta}\,\leq\,\frac{1}{2}\
.
\end{equation}
\item
From~\cite{ben-flore2}, all states $\varrho_I$ with
$N_I=1,2,3,5,7$ are NPT.
Therefore in order for $\varrho_I$ to be PPT, $N_I$ must be equal
to $4$, $6$, $8$ or $9$ or larger.
\item
Since the maximum number of points on
$\mathcal{C}_\alpha\bigcup\mathcal{R}_\beta$ different form
$(\alpha,\beta)$ is at most $7$, all states with $N_I=14,15,16$ are
PPT. Those with $N_I=15,16$ are also separable since $N_I=16$ means
that $\rho_I$ is the maximally mixed state, while, for $N_I=15$,
$\rho_I$ is easily seen to be a separable isotropic
state~\cite{horo99,ben-flore3}.
\item
If the condition in~(\ref{prop1b}) is satisfied for one pair
$(\alpha,\beta)\in L_{16}$, then $N_I=16-6=10$ at the most.
In the following, we shall use an entanglement witness which will allow
us to add to $I$ the site $(\alpha,\beta)$ from a pair
$(\mathcal{C}_\alpha,\mathcal{R}_\beta)$ satisfying~(\ref{prop1b}).
This will open the possibility of checking the entanglement of lattice
states with $N_I=11$.
\end{enumerate}
\bigskip

According to the previous proposition, all $\varrho_I$ not
fulfilling~(\ref{prop1a}) are NPT and thus entangled, while only
some of those which are PPT are recognized as entangled
by~(\ref{prop1b}).

As an example of how the geometric picture works,
consider the following lattice subsets, where the crosses mark those
sites which contribute to $\varrho_I$:
$$
\begin{array}{l                               r}
N_I=8 \quad
 \begin{array}{c|c|c|c|c}
               3 & \quad\! & \quad\! & \quad\! &   \\
               \hline
               2 & \times & \times &  \quad\! & \times  \\
               \hline
               1 & \times & \times & \quad\!  &   \\
               \hline
               0 & \times & \times & \times  &   \\
               \hline
                 & 0 & 1 & 2 & 3
 \end{array} \qquad \qquad \qquad

\qquad
N_I=10 \quad
 \begin{array}{c|c|c|c|c}
               3 & \times & \times & \times  &  \quad\!      \\
               \hline
               2 & \times & \quad\! &  \times &  \quad\!       \\
               \hline
               1 &  \quad\! & \times &  \quad\! & \times       \\
               \hline
               0 & \times & \times  & \times  & \quad\!      \\
               \hline
                 & 0 & 1 & 2 & 3
 \end{array}
\end{array}
$$

$$
\begin{array}{l                               r}
N_I=8 \quad
 \begin{array}{c|c|c|c|c}
               3 & \times & \times & \quad\! & \times  \\
               \hline
               2 & \times & \quad\! &  \times &   \\
               \hline
               1 & \times & \times & \quad\!  & \times  \\
               \hline
               0 & \quad\! & \quad\! & \quad\!  &   \\
               \hline
                 & 0 & 1 & 2 & 3
 \end{array} \qquad \qquad \qquad

\qquad
N_I=10 \quad
 \begin{array}{c|c|c|c|c}
               3 & \times & \times & \quad\!  &  \times      \\
               \hline
               2 & \times & \quad\! &  \times &  \quad\!       \\
               \hline
               1 &  \quad\! & \times &  \quad\! & \times       \\
               \hline
               0 & \times & \times  & \quad\!  & \times      \\
               \hline
                 & 0 & 1 & 2 & 3
 \end{array}
\end{array}
$$

$$
\begin{array}{l                               r}
N_I=10 \quad
 \begin{array}{c|c|c|c|c}
               3 & \quad\! & \times & \times & \times  \\
               \hline
               2 & \times & \times &  \times &  \times \\
               \hline
               1 & \quad\! & \times & \times  & \times  \\
               \hline
               0 & \quad\! & \quad\! & \quad\!  &   \\
               \hline
                 & 0 & 1 & 2 & 3
 \end{array} \qquad \qquad \qquad

\qquad
N_I=11 \quad
 \begin{array}{c|c|c|c|c}
               3 & \times & \times & \times  &  \quad\!      \\
               \hline
               2 & \times & \times &  \times &  \times       \\
               \hline
               1 &  \times & \times &  \times & \quad\!       \\
               \hline
               0 & \quad\! & \quad\!  & \quad\!  & \times      \\
               \hline
                 & 0 & 1 & 2 & 3
 \end{array}
\end{array}
$$

All subsets above identify PPT lattice states according
to~(\ref{prop1a}), but, according to~(\ref{prop1b}), only those in
the left column correspond to (bound) entangled states with certainty.

In the following section, we will develop a criterion strictly stronger
than~(\ref{prop1b}): it will essentially amount to the removal of the
request $(\alpha,\beta)\notin I$ in~(\ref{prop1b}).
This allows us to enrich the class of bound
entangled lattice states, proving that also the states on the right
hand side are entangled. On the other hand, in Section 3.2, we will
further show that some PPT $\varrho_I$ which are not
recognized as entangled by the stronger criterion are indeed separable.

\section{Lattice states and the extended reduction criterion}
\label{lattice_states-Breuer}

We adapt the argument of~\cite{breuer1} and~\cite{hall} to the
lattice states case by extending the reduction
criterion~\cite{horo99} in the following way. We define an
antiunitary map $\vartheta_V$ by its action on a generic $4\times 4$
matrix $B\in M_4(\mathbf{C})$: $\vartheta_V[B] = VB^TV^\dagger$,
where $T$ denotes transposition with respect to a fixed orthonormal
basis and $V:\mathbf{C}^4\mapsto\mathbf{C}^4$ is unitary  and
antisymmetric, $VV^\dagger = V^\dagger V = I_4$, $V^T=-V$.

It follows that, for all $|\psi\rangle\in\mathbf{C}^4$, by expanding
with respect to the chosen basis,
$$
\langle\psi^*|V|\psi\rangle=\sum_{j,k=1}^4\psi_j\,V_{jk}\,\psi_k
=\langle\psi^*|V^T|\psi\rangle=-\langle\psi^*|V|\psi\rangle\ .
$$
Thus, $\vartheta_V[|\psi\rangle\langle\psi|]$ is orthogonal to
$|\psi\rangle\langle\psi|$ so that
$|\psi\rangle\langle\psi|-\vartheta_V[|\psi\rangle\langle\psi|]$ is a
projector for all $|\psi\rangle\in\mathbf{C}^4$, whence the linear map
on $M_4(\mathbf{C})$,
\begin{equation}
\label{reducext}
\Phi_V[B] = \Bigl({\rm Tr}(B)\Bigr)\,I_4\, -\, B\, -\, \vartheta_V[B]\ ,
\end{equation}
preserves positivity.
\bigskip

\noindent \textbf{Remark 2}\quad Since we shall later exhibit PPT
lattice states such that $\displaystyle \Bigl({\rm
id}\otimes\Phi_V\Bigr)[\varrho_I] $ is not positive definite, it
turns out that $\Phi_V$ is  not decomposable, in agreement
with~\cite{hall}.
\bigskip

Notice that the request of antisymmetry and the fact that
$T[\sigma_2]=-\sigma_2$ give the
following expansion of $V$ along the tensor products $\sigma_{\alpha\beta}$
\begin{equation}
\label{V-prop1} V=\sum_{\alpha\neq
2}\,v_{\alpha2}\,\sigma_{\alpha2}\,+\, \sum_{\beta\neq
2}\,v_{2\beta}\,\sigma_{2\beta}\ ,
\end{equation}
where $v_{\alpha2}$ and $v_{2\alpha}$ are complex coefficients
satisfying unitarity constraints.

In order to study the entanglement witnessing power of $\Phi_V$ for the
lattice states, we shall use that $\vartheta_V$ is unitarily
equivalent to the transposition and elaborate on the partial action of the
latter on the projections $P_{\alpha\beta}$ in~(\ref{latt-proj}).
In passing, this will provide a proof of~(\ref{prop1a}) alternative to that
in~\cite{ben-flore3}.

Before proceeding, we make the following useful observations concerning the
algebraic relations among the Pauli matrices and the lattice structure.

The products of two Pauli matrices define four $4\times 4$ hermitian, unitary
matrices $\eta^\alpha$,
\begin{equation}
\label{Pauli-prod}
\sigma_\alpha\sigma_\beta=\sum_{\mu=0}^3\eta^\alpha_{\beta\mu}\,
\sigma_\mu ,
\quad
\eta^\alpha_{\beta\mu}=\frac{1}{2}{\rm Tr}
(\sigma_\alpha\sigma_\beta\sigma_\mu)\ .
\end{equation}
From the explicit expression of the matrix elements and the cyclicity of the
trace operation it follows that
\begin{equation}
\label{Pauli-prod-props}
\eta^\alpha_{\beta\mu}=\eta^\mu_{\alpha\beta}=\eta^\beta_{\mu\alpha}\ ,
\qquad
(\eta^\alpha_{\beta\mu})^*=\eta^\mu_{\beta\alpha}=\eta^\alpha_{\mu\beta}\ .
\end{equation}
The matrices $\eta^\alpha$ are thus hermitian and multiplication
by $\sigma_\alpha$ on the left of both sides of~(\ref{Pauli-prod}) shows that they are
also unitary.
Explicitly, $\eta^0=I_4$ while
\begin{equation}
\label{Pauli-prod2}
\eta^1=\begin{pmatrix}0&1&0&0\cr
1&0&0&0\cr
0&0&0&i\cr
0&0&-i&0\end{pmatrix}\ ,\ \eta^2=\begin{pmatrix}0&0&1&0\cr
0&0&0&-i\cr
1&0&0&0\cr
0&i&0&0\end{pmatrix}\ ,\
\eta^3=\begin{pmatrix}
0&0&0&1\cr
0&0&i&0\cr
0&-i&0&0\cr
1&0&0&0\end{pmatrix}\ .
\end{equation}

Further, given $0\leq\alpha,\beta\leq3$ the Pauli matrix
satisfying~(\ref{Pauli-prod}) is unique, thus any fixed $\alpha$
determines a map $i_\alpha:\{0,1,2,3\}\ni\beta\mapsto
i_\alpha(\beta)\in\{0,1,2,3\}$ defined by
\begin{equation}
\label{Pauli-isom}
\sigma_\alpha\sigma_\beta=\sum_{\mu=0}^3\eta^\alpha_{\beta\mu}\,
\sigma_\mu\,
=\,\eta^\alpha_{\beta i_\alpha(\beta)}\,
\sigma_{i_\alpha(\beta)}\ .
\end{equation}
Multiplying both sides of the above equality by
$\sigma_\alpha$ and taking their hermitian conjugates,
the following useful properties of $i_\alpha$ follow
from~(\ref{Pauli-prod-props}):
\begin{equation}
\label{Pauli-iso-props1}
i_\alpha^2={\rm id}\ ,\quad i_\beta(\alpha)=i_\alpha(\beta)\ .
\end{equation}
Finally, a useful relation involves the matrix $\eta^2$: one can check
that
\begin{equation}
\label{Pauli-iso-props2}
|\eta^2_{\alpha\beta}|=\delta_{\alpha,\beta\oplus 2}\quad
\Longrightarrow\quad i_2(\beta)=\beta\oplus 2\ ,
\end{equation}
where $\oplus$ denotes addition $\hbox{mod}\ 4$.

The action of the partial transposition can now be recast as
\begin{equation}
\label{part-trasp}
\Bigl({\rm id}\otimes T\Bigr)\left[P_{\alpha\beta}\right]=\frac{I_{16}}{4}-
\frac{1}{2}\sum_{(\mu,\nu)\in L_{16}}\Bigl(
\delta_{\alpha,\mu\oplus 2}+\delta_{\beta,\nu\oplus 2}
-2\delta_{\alpha,\mu\oplus 2}\,\delta_{\beta,\nu\oplus 2}\Bigr)\,P_{\mu\nu}\ .
\end{equation}
This can be proved using the following facts:
\begin{enumerate}
\item
by transposing the Pauli matrices one gets
$$
T[\sigma_\alpha]=\sum_{\beta=0}^3\varepsilon_{\alpha\beta}\,\sigma_\beta
\quad \hbox{where}\quad \varepsilon=[\varepsilon_{\mu\nu}]
=\begin{pmatrix}1&0&0&0\cr 0&1&0&0\cr 0&0&-1&0\cr
0&0&0&1\end{pmatrix}\ ;
$$
\item
partially
transposing $P_{00}=|\Psi_+^4\rangle\langle\Psi_+^4|$ yields
the flip $F:A\otimes B\mapsto B\otimes A$, $A,B\in M_4(\mathbf{C})$:
$$
\Bigl({\rm id}\otimes T\Bigl)[P_{00}]
=\frac{1}{4}\sum_{i,j=1}^4|i\rangle\langle
j|\otimes|j\rangle\langle i|=\frac{F}{4}\ ;
$$
\item
$F$ has the spectral decomposition
$F=\sum_{(\alpha,\beta)\in L_{16}}\,\varepsilon_{\alpha\alpha}\,
\varepsilon_{\beta\beta}\, P_{\alpha\beta}$.
\end{enumerate}
Thus one derives
\begin{eqnarray*}
\Bigl({\rm id}\otimes T\Bigr)[P_{\alpha\beta}]&=&
\frac{1}{4}\sum_{(\mu,\nu)\in L_{16}}\,\varepsilon_{\mu\mu}\,
\varepsilon_{\nu\nu}\,
\Bigl(I_4\otimes\sigma_{\alpha\beta}\sigma_{\mu\nu}\Bigr)\,P_{00}\,
\Bigl(I_4\otimes\sigma_{\mu\nu}\sigma_{\alpha\beta}\Bigr)\\
&=&\frac{1}{4}\sum_{(\gamma,\delta)\,,\,(\gamma',\delta')\in L_{16}}\,
(\eta^\alpha\varepsilon\eta^\alpha)_{\gamma'\gamma}\,
(\eta^\beta\varepsilon\eta^\beta)_{\delta\delta'}\,
|\Psi_{\gamma\delta}\rangle\langle\Psi_{\gamma'\delta'}|\ ,
\end{eqnarray*}
whence~(\ref{part-trasp}) follows since,
using~(\ref{Pauli-iso-props2}),
$(\eta^\alpha\varepsilon\eta^\alpha)_{\gamma\gamma'}=\delta_{\gamma,\gamma'}\,
(1-2\delta_{\gamma,\alpha\oplus 2})$.

By means of~(\ref{part-trasp}), the action of the partial transposition
on lattice states yields
\begin{eqnarray}
\label{part-trasp-latt1}
\Bigl({\rm id}\otimes T\Bigr)[\varrho_I]&=&\frac{I_{16}}{4}\,-\,
\frac{1}{2N_I}\sum_{(\mu,\nu)\in L_{16}}\,k^I_{\mu\nu}\, P_{\mu\nu}\quad
\hbox{with}\\
\label{part-trasp-latt2}
k^I_{\mu\nu}&=&\sum_{\alpha\neq \mu\oplus 2}
\chi_I(\alpha,\nu\oplus 2)\,
+\,
\sum_{\beta\neq\nu\oplus 2}\chi_I(\mu\oplus 2,\beta)\ ,
\end{eqnarray}
whence the partial action of the extended reduction map~(\ref{reducext}) gives
\begin{equation}
\label{ext-red-latt}
\Bigl({\rm id}\otimes\Phi_V\Bigr)[\varrho_I]=\frac{1}{2N_I}
\sum_{(\mu,\nu)\in L_{16}}\,k^I_{\mu\nu}\,\Bigl(I_4\otimes V\Bigr)\,
P_{\mu\nu}\,\Bigl(I_4\otimes V^\dagger\Bigr)\,-\,\varrho_I\ .
\end{equation}
\bigskip

From~(\ref{part-trasp-latt1}) and~(\ref{part-trasp-latt2}) the proof
of~(\ref{prop1a}) (and of~(\ref{diag-states-PPT})) easily follows.
Moreover, using~(\ref{ext-red-latt}) one gets:
\bigskip

\noindent
\textbf{Proposition 2}\quad Let $\kappa$ denote the minimum of
$k^I_{\mu\nu}$ in~(\ref{part-trasp-latt2}). Then, a necessary
condition for $\Phi_V$ to reveal the entanglement
of any PPT $\varrho_I$ is $\kappa\leq 1$.
\medskip

\noindent
\textbf{Proof:}\quad
Since unitary transformations do not change an operator's spectrum,
it is more convenient to work with the positive, indecomposable map
$\tilde{\Phi}_V$ defined as follows:
\begin{eqnarray*}
 \Bigl({\rm id} \otimes \widetilde{\Phi}_V\Bigr)[\varrho_I] & = &
\Bigl(I_4 \otimes V^\dagger\Bigr)\Bigl[\Bigl(
{\rm id} \otimes \Phi_V\Bigr)[\varrho_I]\Bigr]\Bigl(I_4 \otimes V\Bigr)\\
& = & -(I_4\!\otimes\!V^\dagger)\varrho_I(I_4\!\otimes\!V) +
\frac{1}{2N_I}\sum_{(\mu,\nu)\in L_{16}}k_{\mu\nu}^I\, P_{\mu\nu}
\label{eqn:Phi_V_tilde}
\end{eqnarray*}
Then, since $\sum_{(\alpha,\beta)\in L_{16}}P_{\alpha\beta}=I_4$,
$$
\rho_I\leq\frac{I_{16}}{N_I}\Longrightarrow
\Bigl({\rm id} \otimes \tilde{\Phi}_V\Bigr)[\varrho_I]
\geq \frac{1}{2N_I}\sum_{(\mu,\nu)\in L_{16}}
(k_{\mu\nu}^I - 2)\, P_{\mu\nu}\ .
$$
\bigskip

From the previous result and~(\ref{part-trasp-latt2}), the extended
reduction criterion based on $\Phi_V$ may detect entangled PPT
lattice states only if

\begin{itemize}
 \item
$ \kappa = 0 $: namely, if there exists at least one pair
$(\mu,\nu)\in L_{16}$ such that $k^I_{\mu\nu}=0$. This is possible
if there exist a column $\mathcal{C}_{\mu\oplus 2}$ and a row
$\mathcal{R}_{\nu\oplus 2}$ which do not contribute to $I$ except,
possibly, for $(\mu\oplus 2,\nu\oplus 2)$;
\item
$ \kappa = 1 $: namely, if there exists at least one pair
$(\mu,\nu)\in L_{16}$ with $k^I_{\mu\nu}=1$. This is possible if
there exist a column $\mathcal{C}_{\mu\oplus 2}$ and a row
$\mathcal{R}_{\nu\oplus 2}$ that contribute to $I$ with only one
element except, possibly, for $(\mu\oplus 2,\nu\oplus 2)$.
\end{itemize}

\subsection{Entangled PPT Lattice States}

In the following, we will consider the second case and postpone the
discussion of the first one to the next section.
We start by observing that if $(\mu\oplus 2,\nu\oplus 2)$ does not
belong to $I$ then, according to~(\ref{prop1b}), the entanglement of
$\rho_I$ is revealed by the positive, indecomposable map devised
in~\cite{ben-flore2}.
We shall show that the extended reduction criterion is able to scoop
the bound entangled $\varrho_I$ in this case, but also when
$(\mu\oplus 2,\nu\oplus 2)$ does belong to $I$, thus being stronger
than the map used in~\cite{ben-flore2}.
\bigskip

\noindent
\textbf{Proposition 3}\quad If $k^I_{\mu\nu}=1$ for
a column $\mathcal{C}_{\mu\oplus 2}$ and a row
$\mathcal{R}_{\nu\oplus 2}$, then
\begin{equation}
\label{prop3}
\langle\Psi_{\mu\nu}\,|\Bigl({\rm id}\otimes
\widetilde{\Phi}_V\Bigr)[\rho_I]\,|\Psi_{\mu\nu}\rangle\,<\, 0\ ,
\end{equation}
independently of whether $(\mu\oplus 2,\nu\oplus 2)$ belongs to $I$ or not.
\medskip

\noindent \textbf{Proof:}\quad Since
$\displaystyle\langle\Psi_+^4|\,A\otimes B\,|\Psi_+^4\rangle ={\rm
Tr}\Bigl(A^T\,B\Bigr)$, using the map~(\ref{Pauli-isom}) one gets
\begin{eqnarray*}
\langle\Psi_{\mu\nu}|\,\Bigl(I_4\otimes V^\dagger\Bigr)\varrho_I
\Bigl(I_4\otimes V\Bigr)\,|\Psi_{\mu\nu}\rangle&=&
\frac{1}{N_I}\sum_{(\alpha,\beta)\in I}\left|\langle\Psi_+^4|\,
I_4\otimes\Bigl(\sigma_{\mu\nu}\,V^\dagger\,\sigma_{\alpha\beta}\Bigr)\,
|\Psi_+^4\rangle\right|^2\\
&=&\frac{1}{4N_I}\sum_{(\alpha,\beta)\in I}\left|{\rm Tr}\Bigl(
\sigma_{i_\mu(\alpha)i_\nu(\beta)}\,V\Bigr)\right|^2\\
&=&\frac{1}{N_I}\sum_{(\alpha,\beta)\in I}\,
\left|v_{i_\mu(\alpha)\,i_\nu(\beta)}\right|^2\ ,
\end{eqnarray*}
where, due to~(\ref{V-prop1}), either $i_\mu(\alpha)=0,1,3$ with
$i_\nu(\beta)=2$ or $i_\nu(\beta)=0,1,3$ with $i_\mu(\alpha)=2$.

Because of~(\ref{Pauli-iso-props1}) and~(\ref{Pauli-iso-props2}), the
first case corresponds to $\beta=\nu\oplus 2$ and
$\alpha\neq\mu\oplus 2$ and the second one to $\alpha=\mu\oplus 2$ and
$\beta\neq\nu\oplus 2$, whence
\begin{eqnarray*}
&&
\langle\Psi_{\mu\nu}|\,\Bigl(I_4\otimes V^\dagger\Bigr)\varrho_I
\Bigl(I_4\otimes V\Bigr)\,|\Psi_{\mu\nu}\rangle=
\frac{1}{N_I}\left\{
\sum_{\alpha\neq\mu\oplus 2}\chi_I(\alpha,\nu\oplus 2)\,
\left|v_{i_\mu(\alpha)\,2}\right|^2\right.\\
&&\hskip 2cm
\,+\,\left.
\sum_{\beta\neq\nu\oplus 2}\chi_I(\mu\oplus 2,\beta)\,
\left|v_{2\,i_\nu(\beta)}\right|^2\right\}\ ,
\end{eqnarray*}
whence
\begin{eqnarray*}
&&
\langle\Psi_{\mu\nu}|\,\Bigl({\rm id}\otimes\widetilde{\Phi}_V\Bigr)
[\varrho_I]\,|\Psi_{\mu\nu}\rangle=
\frac{1}{2N_I}\left\{
\sum_{\alpha\neq\mu\oplus 2}\chi_I(\alpha,\nu\oplus 2)\,
\Bigl(1-2\left|v_{i_\mu(\alpha)\,2}\right|^2\Bigr)\right.\\
&&\hskip 2cm
\,+\,\left.
\sum_{\beta\neq\nu\oplus 2}\chi_I(\mu\oplus 2,\beta)\,
\Bigl(1-2\left|v_{2\,i_\nu(\beta)}\right|^2\Bigr)\right\}\ .
\end{eqnarray*}
Suppose that the point of $I$ contributing to $k^I_{\mu\nu}=1$ be
$(\alpha,\nu\oplus 2)$ on the row $\mathcal{R}_{\nu\oplus 2}$, then
$V=\sigma_{i_\mu(\alpha)2}$ yields the result.
A similar argument holds if the contributing point is
$(\mu\oplus 2,\beta)$ on the column $\mathcal{C}_{\mu\oplus 2}$.
\bigskip

\noindent
\textbf{Remark 3}\quad
The previous result is not sensitive to whether the point
$(\mu\oplus 2,\nu\oplus 2)$ belonging to the column
$\mathcal{C}_{\mu\oplus 2}$ and row $\mathcal{R}_{\nu\oplus2}$ does
contribute to $I$ or not.
If it does not, the extended reduction criterion reveals all the
bound entangled lattice states already revealed by the criterion
in~\cite{ben-flore2}; if it does belong to $I$, new bound entangled
states not revealed by the latter are seen by the extended reduction
map.
\bigskip

\subsection{Separable PPT Lattice States}

By direct inspection, the argument of the previous proposition, by
which all bound entangled lattice states with $\kappa=1$ are
detectable by the extended reduction criterion, cannot be applied
when $\kappa=0$.

It turns out that the case $\kappa=0=k^I_{\mu\nu}$ for some $(\mu,\nu)\in
L_{16}$ with $(\mu\oplus 2,\nu\oplus 2)\notin I$ can be studied by
other means.
The following observation turns out to be useful.
\bigskip

\noindent
\textbf{Remark 4}
As observed in~\cite{ben-flore2}, one can operate local unitary
transformations transforming lattice states into lattice states
without changing either their cardinality $N_I$ or their entanglement
properties.
Therefore, one can actually subdivide the lattice states into
equivalence classes by operating on them with local unitaries whose
only effect is to map the set of subsets $I\in L_{16}$ of a same
cardinality into itself.
Indeed, given two $4\times 4$ matrices $U,V$, the structure of
$\vert\Psi_+^4\rangle$ is such that
$$
U\otimes V|\Psi_{\alpha\beta}\rangle
=I_4\otimes V\sigma_{\alpha\beta}U^T|\Psi_+^4\rangle\ .
$$
We distinguish the following cases:
\begin{enumerate}
\item
Choosing $U=I_4$ and $V=\sigma_{\alpha\beta}$, we get
$I_4\otimes\sigma_{\alpha\beta}P_{\alpha\beta}I_4\otimes\sigma_{\alpha\beta}=
P_{00}$, so that we can move any chosen column $\mathcal{C}_\alpha$
and row $\mathcal{R}_\beta$ into $\mathcal{C}_0$ and
$\mathcal{R}_0$.
\item
Let $U=U_1\otimes U_2$ and $V=U_1^*\otimes U_2^*$, where $U_1$,
respectively $U_2$ are two unitary $2\times 2$ matrices that rotate
the Pauli matrices $\sigma_i$, $\sigma_{i'}$, respectively
$\sigma_j$, $\sigma_{j'}$, one into the other (apart possibly from a
$-$ sign) keeping fixed the third one, $\sigma_{i''}$, respectively
$\sigma_{j''}$ (and of course $\sigma_0$):
\begin{eqnarray*}
&&
U_1\sigma_iU^\dagger_1=\pm\sigma_{i'}\ ,\quad
U_1\sigma_{i'}U_1^\dagger
=\mp\sigma_i\ ,\quad i,i'\neq0\qquad\hbox{and}\\
&&
U_2\sigma_jU^\dagger_2=\pm\sigma_{j'}\ ,\quad
U_2\sigma_{j'}U_2^\dagger=\mp\sigma_j\ ,\quad j,j'\neq0\ .
\end{eqnarray*}
Then, any two columns $\mathcal{C}_i$, $\mathcal{C}_{i'}$,
respectively rows $\mathcal{R}_j$, $\mathcal{R}_{j'}$ can be changed
one into the other,
leaving fixed the columns $\mathcal{C}_0$, $\mathcal{C}_{i''}$,
respectively the rows $\mathcal{R}_{0}$, $\mathcal{R}_{j''}$:
\begin{eqnarray*}
V\otimes U\,P_{ij}\,
V^\dagger\otimes U^\dagger&=&
I_4\otimes(U_1\sigma_iU_1^\dagger\otimes U_2\sigma_j
U_2^\dagger)\,P_{00}\,I_4\otimes(U_1\sigma_iU_1^\dagger\otimes U_2\sigma_j
U_2^\dagger)\\
&=&
I_4\otimes\sigma_{i'j'}\,P_{00}\,I_4\otimes\sigma_{i'j'}=P_{i'j'}\ .
\end{eqnarray*}
\item
Finally, choosing $U=I_4$ and $V$ the flip operator
($V\vert\Psi_+^4\rangle=\vert\Psi_+^4\rangle$,
$V\sigma_{\alpha\beta}V=\sigma_{\beta\alpha}$), one gets:
$$
I_4\otimes V\,P_{\alpha\beta}\,I_4\otimes V=
I_4\otimes(V\sigma_{\alpha\beta}V)
\,|\Psi^4_+\rangle\langle\Psi_+^4|\,I_4\otimes(V\sigma_{\alpha\beta}V)
=P_{\beta\alpha}\ .
$$
One can thus exchange columns and rows of a subset $I\subseteq
L_{16}$ obtaining a new lattice state in the same equivalence class,
hence with the same entanglement properties.
\end{enumerate}
\medskip

The previous considerations allow us to prove the following result.
\medskip

\noindent
\textbf{Proposition 4}\quad
If $k^I_{\mu\nu}=0$ and $(\mu\oplus2,\nu\oplus2)$ does not belong to
\textit{I}, all the corresponding PPT states $\varrho_I$ are  separable.
\medskip

\noindent \textbf{Proof:}\quad The hypothesis means that the column
$\mathcal{C}_{\mu\oplus 2}$ and the row $\mathcal{R}_{\nu\oplus 2}$
do not contribute to $I$. Thus $7$ points on the square lattice
$L_{16}$ must be excluded and at the most $9$ elements of $L_{16}$
label $\varrho_I$. From Remark 1.2 we know that all states
$\varrho_I$ with $N_I=1,2,3,5,7$ are NPT. Therefore in order for
$\varrho_I$ to be PPT, $N_I$ must be equal to $4$, $6$, $8$ or $9$.
From Remark 1.4, we also know that all PPT $\varrho_I$ with $N_I=4$
are separable, in particular those associated with the following
graphs (and all those in their equivalence classes as
in Remarks 5)
$$
\begin{array}{l                               r}
 \begin{array}{c|c|c|c|c}
               3 & \quad\! & \quad\! & \quad\! & \times  \\
               \hline
               2 & \quad\! & \quad\! &  \times &   \\
               \hline
               1 & \quad\! & \times & \quad\!  &   \\
               \hline
               0 & \times & \quad\! & \quad\!  &   \\
               \hline
                 & 0 & 1 & 2 & 3
 \end{array} \hskip 1cm\ ,\hskip 1cm

\qquad
 \begin{array}{c|c|c|c|c}
               3 & \quad\! & \quad\! & \quad\!  &  \quad\!      \\
               \hline
               2 & \quad\! & \quad\! &  \quad\! &  \quad\!       \\
               \hline
               1 &  \times & \times &  \quad\! & \quad\!       \\
               \hline
               0 & \times & \times  & \quad\!  & \quad\!      \\
               \hline
                 & 0 & 1 & 2 & 3
 \end{array}
\end{array}\quad .
$$

In the following, we shall use these separable rank-$4$ lattice
states to write other lattice states of higher rank as their convex
combinations which will thus also result separable; for sake of
simplicity, we shall identify graphs with lattice states and the
convex decompositions with weighted sums of graphs.

Because of Remark 4, we can always perform a unitary transformation
so that the column and row which are assumed to be completely empty
correspond to $\mathcal{C}_0$ and $\mathcal{R}_0$.
\vfill\break

\noindent $\bullet$\quad Case $N_I=9$

\noindent
We have only one possible equivalence class represented by
the state $\varrho^9_I$ associated with the graph
$$ \varrho_I^9 : \qquad
 \begin{array}{c|c|c|c|c}
               3 & \quad\! & \times  & \times &  \times    \\
               \hline
               2 & \quad\!  & \times & \times & \times     \\
               \hline
               1 &  \quad\! & \times & \times & \times     \\
               \hline
               0 & \quad\! & \quad\!  & \quad\! & \quad\!     \\
               \hline
                 & 0 & 1 & 2 & 3
 \end{array} \quad .
$$

This lattice state can be convexly decomposed in terms of $9$ rank-$4$
PPT (hence separable) lattice states as follows:

$$
\begin{array}{l                    c                   r}
 \quad \frac{1}{9} \quad
 \begin{array}{c|c|c|c|c}
               3 & \quad\!  &  \quad\! &  \times &  \times    \\
               \hline
               2 &  \quad\! & \quad\! & \times & \times     \\
               \hline
               1 &  \quad\! & \quad\! & \quad\! & \quad\!     \\
               \hline
               0 & \quad\! & \quad\!  & \quad\! & \quad\!     \\
               \hline
                 & 0 & 1 & 2 & 3
 \end{array} \qquad
$$

$$
 +\; \frac{1}{9} \quad
 \begin{array}{c|c|c|c|c}
               3 & \quad\!  &  \quad\! &  \quad\! &  \quad\!    \\
               \hline
               2 &  \quad\! & \times & \times & \quad\!     \\
               \hline
               1 &  \quad\! & \times & \times & \quad\!     \\
               \hline
               0 & \quad\! & \quad\!  & \quad\! & \quad\!     \\
               \hline
                 & 0 & 1 & 2 & 3
 \end{array} \qquad
$$

$$
 +\; \frac{1}{9} \quad
 \begin{array}{c|c|c|c|c}
               3 & \quad\!  &  \times &  \times &  \quad\!    \\
               \hline
               2 &  \quad\! & \times & \times & \quad\!     \\
               \hline
               1 &  \quad\! & \quad\! & \quad\! & \quad\!     \\
               \hline
               0 & \quad\! & \quad\!  & \quad\! & \quad\!     \\
               \hline
                 & 0 & 1 & 2 & 3
 \end{array} \qquad
$$
\end{array}
$$

$$
\begin{array}{l                 c                  r}
 +\; \frac{1}{9} \quad
 \begin{array}{c|c|c|c|c}
               3 & \quad\!  &  \times &  \times &  \quad\!    \\
               \hline
               2 &  \quad\! & \quad\! & \quad\! & \quad\!     \\
               \hline
               1 &  \quad\! & \times & \times & \quad\!     \\
               \hline
               0 & \quad\! & \quad\!  & \quad\! & \quad\!     \\
               \hline
                 & 0 & 1 & 2 & 3
 \end{array} \qquad
$$

$$
+\; \frac{1}{9} \quad
 \begin{array}{c|c|c|c|c}
               3 & \quad\!  &  \quad\! &  \quad\! &  \quad\!    \\
               \hline
               2 &  \quad\! & \times & \quad\! & \times     \\
               \hline
               1 &  \quad\! & \times & \quad\! & \times     \\
               \hline
               0 & \quad\! & \quad\!  & \quad\! & \quad\!     \\
               \hline
                 & 0 & 1 & 2 & 3
 \end{array} \qquad
$$

$$
+\; \frac{1}{9} \quad
 \begin{array}{c|c|c|c|c}
               3 & \quad\!  &  \times &  \quad\! &  \times    \\
               \hline
               2 &  \quad\! & \quad\! & \quad\! & \quad\!     \\
               \hline
               1 &  \quad\! & \times & \quad\! & \times     \\
               \hline
               0 & \quad\! & \quad\!  & \quad\! & \quad\!     \\
               \hline
                 & 0 & 1 & 2 & 3
 \end{array} \qquad
$$
\end{array}
$$

$$
\begin{array}{l                     c                     r}
 +\; \frac{1}{9} \quad
 \begin{array}{c|c|c|c|c}
               3 & \quad\!  &  \quad\! &  \quad\! &  \quad\!    \\
               \hline
               2 &  \quad\! & \quad\! & \times & \times     \\
               \hline
               1 &  \quad\! & \quad\! & \times & \times     \\
               \hline
               0 & \quad\! & \quad\!  & \quad\! & \quad\!     \\
               \hline
                 & 0 & 1 & 2 & 3
 \end{array} \qquad
$$

$$
+\; \frac{1}{9} \quad
 \begin{array}{c|c|c|c|c}
               3 & \quad\!  &  \quad\! &  \times &  \times    \\
               \hline
               2 &  \quad\! & \quad\! & \quad\! & \quad\!     \\
               \hline
               1 &  \quad\! & \quad\! & \times & \times     \\
               \hline
               0 & \quad\! & \quad\!  & \quad\! & \quad\!     \\
               \hline
                 & 0 & 1 & 2 & 3
 \end{array} \qquad
$$

$$
+\; \frac{1}{9} \quad
 \begin{array}{c|c|c|c|c}
               3 & \quad\!  &  \times &  \quad\! &  \times    \\
               \hline
               2 &  \quad\! & \times & \quad\! & \times     \\
               \hline
               1 &  \quad\! & \quad\! & \quad\! & \quad\!     \\
               \hline
               0 & \quad\! & \quad\!  & \quad\! & \quad\!     \\
               \hline
                 & 0 & 1 & 2 & 3
 \end{array} \qquad
$$
\end{array}
$$
In decomposing $\varrho_I^9$ as done above, each contributing site
from the decomposers appears the same number of times ($4$ in this
case) and each decomposer contributes with equal weight ($1/9$ in
this case); then, since the decomposers have $N_I=4$, each
contributing site appears with weight $4\times 1/36=1/9$ in the
resulting convex decomposition. From Proposition 1 we know the state
$\varrho_I^9$ is PPT and, as a convex combination of separable
states, it is separable.
\medskip

\noindent $\bullet$\quad Case $N_I=8$

\noindent
With $\mathcal{C}_0$ and $\mathcal{R}_0$ completely empty
there are nine possible states, which can be obtained directly from
the above state $\varrho_I^9$ with $N_I=9$ by eliminating one
element, i.e.
$$
\begin{array}{l                     c                     r}
 \begin{array}{c|c|c|c|c}
               3 &  \quad\!   & \times  &  \times & \times     \\
               \hline
               2 &  \quad\! &  \times & \times & \times     \\
               \hline
               1 & \quad\!  & \quad\! & \times & \times     \\
               \hline
               0 & \quad\! & \quad\!  & \quad\! &      \\
               \hline
                 & 0 & 1 & 2 & 3
 \end{array} \quad\ ,\quad
$$

$$
 \begin{array}{c|c|c|c|c}
               3 &  \quad\!   & \times  &  \times & \times     \\
               \hline
               2 &  \quad\! &  \times & \times & \times     \\
               \hline
               1 & \quad\!  & \times & \quad\! & \times     \\
               \hline
               0 & \quad\! & \quad\!  & \quad\! &      \\
               \hline
                 & 0 & 1 & 2 & 3
 \end{array} \quad\ ,\quad
$$

$$
 \begin{array}{c|c|c|c|c}
               3 &  \quad\!   & \times  &  \times & \times     \\
               \hline
               2 &  \quad\! &  \times & \times & \times     \\
               \hline
               1 & \quad\!  & \times & \times & \quad\!     \\
               \hline
               0 & \quad\! & \quad\!  & \quad\! &      \\
               \hline
                 & 0 & 1 & 2 & 3
 \end{array} \qquad etc.
$$
\end{array}
$$

By exchanging columns and rows as explained in Remark 5,
all of these nine possible states are unitarily equivalent to the state
$$ \varrho_I^8 : \qquad
 \begin{array}{c|c|c|c|c}
               3 & \quad\! & \times  & \times &  \times    \\
               \hline
               2 & \quad\!  & \times & \quad\! & \times     \\
               \hline
               1 &  \quad\! & \times & \times & \times     \\
               \hline
               0 & \quad\! & \quad\!  & \quad\! & \quad\!     \\
               \hline
                 & 0 & 1 & 2 & 3
 \end{array} \qquad
$$

Therefore in order to study these lattice states, it is sufficient
to see whether $\varrho_I^8$ is separable or not. From Proposition 1
we know $\varrho_I^8$ is PPT; moreover, it can be convexly
decomposed as follows:
$$
 \begin{array}{l                                              r}
 \begin{array}{c|c|c|c|c}
               3 & \quad\!  & \times & \times  & \times     \\
               \hline
               2 & \quad\!  &  \times   &  \quad\!  &  \times   \\
               \hline
               1 &  \quad\! & \times &  \times & \times     \\
               \hline
               0 &    &     &     &      \\
               \hline
                 & 0 & 1 & 2 & 3
 \end{array}  \qquad = $$

$$

 \quad \frac{1}{4} \quad
 \begin{array}{c|c|c|c|c}
               3 &  \quad\!  &  \times  &  \times   &      \\
               \hline
               2 &  \quad\!  &  \quad\!  & \quad\!  &      \\
               \hline
               1 &  \quad\! & \times &  \times   &       \\
               \hline
               0 & \quad\!  &  \quad\!   &  \quad\! &      \\
               \hline
                 & 0 & 1 & 2 & 3
 \end{array} \qquad
 \end{array}
$$

$$
\begin{array}{l                      c                       r}
 + \; \frac{1}{4} \quad
 \begin{array}{c|c|c|c|c}
               3 &  \quad\!  &  \quad\!  & \quad\!   &      \\
               \hline
               2 &  \quad\! & \times &  \quad\!   & \times     \\
               \hline
               1 &  \quad\! & \times &  \quad\!   & \times     \\
               \hline
               0 &  \quad\!  & \quad\!   &  \quad\!   &      \\
               \hline
                 & 0 & 1 & 2 & 3
 \end{array} \quad
  $$

 +\; \frac{1}{4} \quad
 \begin{array}{c|c|c|c|c}
               3 &  \quad\!  &  \quad\!  &  \times &  \times    \\
               \hline
               2 & \quad\!  & \quad\! & \quad\!  &      \\
               \hline
               1 &  \quad\!  &  \quad\!  &  \times   &  \times     \\
               \hline
               0 & \quad\! & \quad\!  & \quad\! &      \\
               \hline
                 & 0 & 1 & 2 & 3
 \end{array} \quad $$

$$ +\; \frac{1}{4} \quad
 \begin{array}{c|c|c|c|c}
               3 & \quad\!   & \times   &  \quad\! &  \times    \\
               \hline
               2 & \quad\!  & \times & \quad\!  &  \times    \\
               \hline
               1 &  \quad\!  & \quad\!   & \quad\! &       \\
               \hline
               0 & \quad\! & \quad\!  & \quad\! &      \\
               \hline
                 & 0 & 1 & 2 & 3
 \end{array}
 \end{array}
$$
In the above decomposition of $\varrho_I^8$ each contributing site
appears with weight $1/4$ in two different PPT decomposers of
rank-$4$;
therefore, in the resulting convex combination its weight is
$2\times 1/16=1/8$. Again, $\varrho_I^8$ is a mixture of separable
states and thus separable itself.
\medskip

\noindent $\bullet$\quad Case $N_I=6$

\noindent
We have to fit six elements into nine points of the square
lattice $L_{16}$ (excluding row $\mathcal{R}_0$ and column
$\mathcal{C}_0$). According to Proposition 1, in order for the
states to be PPT there cannot be more than two elements in each of
the three free columns or rows, i.e. all states such as the following are
NPT:
$$
\begin{array}{l                              r}
 \begin{array}{c|c|c|c|c}
               3 & \quad\! & \times & \times &   \\
               \hline
               2 & \quad\! & \times &  \quad\! &   \\
               \hline
               1 & \quad\! & \times & \times  & \times  \\
               \hline
               0 & \quad\! & \quad\! & \quad\!  &   \\
               \hline
                 & 0 & 1 & 2 & 3
 \end{array}
\quad\ ,\quad
 \begin{array}{c|c|c|c|c}
               3 & \quad\! & \times & \quad\! & \quad\!  \\
               \hline
               2 & \quad\! & \times &  \times & \times  \\
               \hline
               1 & \quad\! & \times & \times  & \quad\!  \\
               \hline
               0 & \quad\! & \quad\! & \quad\!  &   \\
               \hline
                 & 0 & 1 & 2 & 3
 \end{array}
\end{array}\quad .
$$

From Proposition 1 it also follows that all states with only two
elements in each row (column) but no empty column (row) except for
$\mathcal{C}_0$ ($\mathcal{R}_0$) are NPT, as, for instance, the following
ones:
$$
\begin{array}{l                              r}
\begin{array}{c|c|c|c|c}
               3 & \quad\! & \times & \times &   \\
               \hline
               2 & \quad\! & \times &  \quad\! & \times  \\
               \hline
               1 & \quad\! & \times & \times  &   \\
               \hline
               0 & \quad\! & \quad\! & \quad\!  &   \\
               \hline
                 & 0 & 1 & 2 & 3
\end{array}
\quad\ ,\quad
 \begin{array}{c|c|c|c|c}
               3 & \quad\! & \quad\! & \quad\! & \times  \\
               \hline
               2 & \quad\! & \times &  \times & \quad\!  \\
               \hline
               1 & \quad\! & \times & \times  & \times  \\
               \hline
               0 & \quad\! & \quad\! & \quad\!  &   \\
               \hline
                 & 0 & 1 & 2 & 3
\end{array}
\end{array}\quad .
$$

So the only possible way to construct PPT states $\varrho_I$
with $N_I=6$ and the row $\mathcal{R}_0$ and column $\mathcal{C}_0$ completely
empty, is by putting two elements in each row (column) and leaving
another column (row) empty, i.e. lattice states such as
$
\varrho_I^6=\begin{array}{c|c|c|c|c}
               3 & \quad\! & \times & \times &   \\
               \hline
               2 & \quad\! & \times &  \times &   \\
               \hline
               1 & \quad\! & \times & \times  &   \\
               \hline
               0 & \quad\! & \quad\! & \quad\!  &   \\
               \hline
                 & 0 & 1 & 2 & 3
\end{array}
$
and all those in its equivalence class obtained by exchanging columns
(rows) among themselves by unitary rotations and row into columns by
means of the flip operator as described in Remark 5.

This state was already showed to be separable in~\cite{ben-flore2}, where an
explicit decomposition had to be worked out; using the strategy of above,
the proof is now much simpler.
Indeed, the state $\varrho_I^6$ can be written as follows:
$$
\varrho_I^6\ =\  \frac{1}{3} \                                                   \begin{array}{c|c|c|c|c}
               3 & \quad\! &  \quad\!  &  \quad\!   &      \\
               \hline
               2 & \quad\! &  \times  &  \times   &      \\
               \hline
               1 &  \quad\! & \times &  \times   &       \\
               \hline
               0 & \quad\! & \quad\!  &  \quad\!   &      \\
               \hline
                 & 0 & 1 & 2 & 3
 \end{array}
\ + \ \frac{1}{3}\
\begin{array}{c|c|c|c|c}
               3 & \quad\!  & \times & \times &      \\
               \hline
               2 &  \quad\! & \times & \times  &      \\
               \hline
               1 &  \quad\! & \quad\! &  \quad\!  &       \\
               \hline
               0 & \quad\! & \quad\! & \quad\!  &      \\
               \hline
                 & 0 & 1 & 2 & 3
 \end{array} \ +\
\frac{1}{3}\
 \begin{array}{c|c|c|c|c}
               3 & \quad\! & \times & \times &      \\
               \hline
               2 & \quad\!  & \quad\! & \quad\!  &      \\
               \hline
               1 & \quad\! & \times & \times &       \\
               \hline
               0 & \quad\! & \quad\!  &  \quad\! &      \\
               \hline
                 & 0 & 1 & 2 & 3
\end{array}\quad .
$$
In the above decomposition of $\varrho_I^6$ each element appears in
two different rank-$4$ lattice states, so that its total weight in the
resulting decomposition is $2\times 1/12=1/6$.
Thus, since $\varrho_I^6$ is a convex combination of PPT
states with $N_I=4$, it is separable.
Therefore all PPT lattice states with $N_I=6$ and a column and row
which are completely empty are separable because they are unitarily equivalent
to $\varrho_I^6$.

\subsection{Lattice States Classification}

From Remarks 1, the only lattice states whose entanglement structure
is not under control are those with $N_I=6$ and $8\leq N_I\leq14$.
We now show that the case $N_I=6$ can also be completely controlled.

As in the proof of Proposition 4, the clue is Remark 4: lattice
states can be grouped into equivalence classes, each element in a
class being obtainable from any of its partners in that class by
the local action of unitaries. Members of a class share the same
entanglement properties.

This allows us to consider without restriction those graphs where the
largest number of sites, that we shall denote by $n_0$, contributing to
$I$ by a column are in $\mathcal{C}_0$ from bottom to top.
We can thus proceed by distinguishing various cases.
\medskip

\noindent
$\bullet$\quad Case $n_0=4$

\noindent
The remaining two contributing sites will necessarily be located on
some of the other columns thus violating condition~(\ref{prop1a}):
all lattice states with $n_0=4$ are thus NPT-entangled.
\medskip

\noindent
$\bullet$\quad Case $n_0=3$

\begin{enumerate}
\item
If any of the sites of $\mathcal{R}_3$ contribute to $I$, then the
corresponding lattice state violates~(\ref{prop1a}) and is
NPT-entangled.
\item
If $\mathcal{R}_3\bigcap I$ is empty and one of the rows
$\mathcal{R}_{0,1,2}$ contributes with more than $1$
site to $I$,~(\ref{prop1a}) is again violated and the lattice state is
NPT-entangled.
\item
If $\mathcal{R}_3\bigcap I$ is empty and the $3$ rows
$\mathcal{R}_{0,1,2}$ contribute with only $1$ site
to $I$ and there is a column, say $\mathcal{C}_\alpha$, $\alpha\neq0$,
with only $1$ site contributing to $I$, then the sufficient condition
in Proposition 1 applies to $\mathcal{C}_\alpha$ and $\mathcal{R}_3$.
All these lattice states are thus PPT entangled.
\item
If $\mathcal{R}_3\bigcap I$ is empty and there is a column
(which we can always suppose to be $\mathcal{C}_1$) with $2$ elements
in $I$, the remaining contributing site
makes these states violate the PPTness condition in Proposition 1.
These lattice states are thus NPT-entangled.
In fact, the row passing through the last contributing site, let it be
$\mathcal{R}_\beta$, intersects
$\mathcal{C}_0$ in one site and $\mathcal{C}_1$ in none, whence
$(\mathcal{C}_1\bigcup\mathcal{R}_\beta)\bigcap I$ contains $4$ sites.
\item
If $\mathcal{R}_3\bigcap I$ is empty and the $3$ sites of $I$ from the
rows $\mathcal{R}_{0,1,2}$
are on a same column $\mathcal{C}_\alpha$, $\alpha\neq0$, then we are
in the situation of Proposition 4. These lattice states are thus separable.
\end{enumerate}
\medskip

\noindent
$\bullet$\quad Case $n_0=2$

\begin{enumerate}
\item
If one of the rows $\mathcal{R}_{2,3}$ contributes with more than $1$
site to $I$, condition~(\ref{prop1a}) is violated and the state $\varrho_I$
is NPT-entangled.
\item
If the rows  $\mathcal{R}_{2,3}$ altogether contribute to $I$ with $1$ site at
the most, then there will be at least one of the rows
$\mathcal{R}_{0,1}$ intercepting $I$ in not less than three sites.
In this case, using the flip operator as in Remark 4.3,
rows can be turned into columns and we are back with either $n_0=3$ or
$n_0=4$.
\item
If each of the rows $\mathcal{R}_{2,3}$ intersects $I$ in only one point
and these two elements lie on a same column, again~(\ref{prop1a}) is
violated and
$\varrho_I$ is NPT-entangled.
Indeed, the two elements can always be thought to be along the column
$\mathcal{C}_1$ which cannot then contain other sites from $I$.
Therefore,~(\ref{prop1a}) is violated by any remaining contributing
site exactly as in point $4$ when $n_0=3$.
The following is a representative of such a possibility:
$$
 \begin{array}{c|c|c|c|c}
               3 & \quad\! & \times & \quad\! &   \\
               \hline
               2 & \quad\! & \times &  \quad\! &  \\
               \hline
               1 & \times & \quad\!& \quad\!  &\times   \\
               \hline
               0 & \times & \quad\! & \times  &   \\
               \hline
                 & 0 & 1 & 2 & 3
\end{array}\quad .
$$
\item
The only remaining case is thus represented by $
\begin{array}{c|c|c|c|c}
               3 & \quad\! & \quad\! & \quad\! &\times   \\
               \hline
               2 & \quad\! & \quad\! &  \times &  \\
               \hline
               1 & \times & \times& \quad\!  &   \\
               \hline
               0 & \times & \times\! & \quad\!  &   \\
               \hline
                 & 0 & 1 & 2 & 3
 \end{array}\quad .
$
This state is separable as it can be decomposed into $3$ PPT
lattice states $\rho_I$ with $N_I=4$. As in Proposition
$4$, we indicate the weights with which the various graphs contribute
to the decomposition:
\begin{eqnarray*}
\begin{array}{c|c|c|c|c}
               3 & \quad\! & \quad\! & \quad\! &\times   \\
               \hline
               2 & \quad\! & \quad\! &  \times &  \\
               \hline
               1 & \times & \times& \quad\!  &   \\
               \hline
               0 & \times & \times\! & \quad\!  &   \\
               \hline
                 & 0 & 1 & 2 & 3
\end{array}&=&\frac{1}{3}\quad
\begin{array}{c|c|c|c|c}
               3 & \quad\! & \quad\! & \quad\! &\times   \\
               \hline
               2 & \quad\! & \quad\! &  \times &  \\
               \hline
               1 & \quad\!& \times& \quad\!  &   \\
               \hline
               0 & \times & \quad\! & \quad\!  &   \\
               \hline
                 & 0 & 1 & 2 & 3
 \end{array}
\quad+\quad\frac{1}{3}\quad
\begin{array}{c|c|c|c|c}
               3 & \quad\! & \quad\! & \quad\! &   \\
               \hline
               2 & \quad\! & \quad\! &  \quad\! &  \\
               \hline
               1 & \times & \times& \quad\!  &   \\
               \hline
               0 & \times & \times\! & \quad\!  &   \\
               \hline
                 & 0 & 1 & 2 & 3
 \end{array}\\
&+&\frac{1}{3}\quad
\begin{array}{c|c|c|c|c}
               3 & \quad\! & \quad\! & \quad\! &\times   \\
               \hline
               2 & \quad\! & \quad\! &  \times &  \\
               \hline
               1 & \times & \quad\!& \quad\!  &   \\
               \hline
               0 & \quad\! & \times\! & \quad\!  &   \\
               \hline
                 & 0 & 1 & 2 & 3
\end{array}\quad .
\end{eqnarray*}
\end{enumerate}

\section{Conclusions}
\label{conclu}

Because of the difficulty in finding appropriate entanglement
witnesses, the detection of bipartite entanglement and its
qualification as bound or free still represents a challenge from the
point of view of  mathematical physics. Of course, the issue at
stake is the lack of a structural characterization of positive, but
not completely positive maps.

In order to deepen our actual understanding of positivity of linear
maps, in particular with respect to their indecomposability and the
corresponding phenomenon of bound entanglement, it is still
important to provide classes of states together with maps that are
able to detect the different aspects of their entanglement.

To this end, in this paper we studied a class of bipartite states,
called lattice states, each party consisting of $2$-qubit systems,
which are nicely associated with subsets $I$ of points on a square
lattice with $16$ elements.
By using a suitably adapted extended reduction criterion,
we exposed certain relations between bound entanglement and the
geometric pattern of the subsets $I$ labeling the lattice states.
More precisely, based on an existing complete characterization of which
of the lattice states are PPT, we showed that

\begin{enumerate}
\item
all PPT lattice states labeled by a subset $I$ such that there exist
a column $\mathcal{C}_\alpha$ and a row $\mathcal{R}_\beta$ of the
lattice contributing to $I$ with only one point different from
$(\alpha,\beta)$ are entangled.
\item
If $(\alpha,\beta)\notin I$, the extended reduction criterion reproduces
a result already obtained with a different indecomposable positive
map.
If $(\alpha,\beta)\in I$, new bound entangled lattice states are
witnessed.
\item
All PPT lattice states labeled by a subset $I$ that entirely
excludes a column and a row are separable.
\end{enumerate}

However, a thorough classification of the class of lattice states
still requires stronger criteria than the ones presented in this
paper. Indeed, the positive map $\Phi_V$~(\ref{reducext}) is
unfortunately not an exhaustive entanglement witness; for instance,
it is indecisive in the following cases of PPT lattice states. The
first three examples are with $\kappa=0$ for some $(\alpha,\beta)\in
I$, while the last one, with $N_I=11$, is a geometric pattern which
cannot be controlled by any of the methods employed in this paper:
$$
N_I=8:\quad \begin{array}{c|c|c|c|c}
               3 & \quad\! & \times & \times &\times   \\
               \hline
               2 & \quad\! & \times &  \quad\! &\times  \\
               \hline
               1 & \quad\! & \times& \quad\!  &\times   \\
               \hline
               0 & \times & \quad\! & \quad\!  &   \\
               \hline
                 & 0 & 1 & 2 & 3
\end{array}
\ ,\quad\,
N_I=9:\quad \begin{array}{c|c|c|c|c}
               3 & \quad\! & \times & \times &\times   \\
               \hline
               2 & \quad\! & \times &\quad\! &\times  \\
               \hline
               1 & \quad & \times & \times  &\times   \\
               \hline
               0 & \times & \quad\! & \quad\!  &   \\
               \hline
                 & 0 & 1 & 2 & 3
 \end{array}\quad ,
$$
$$
N_I=10:\quad
\begin{array}{c|c|c|c|c}
               3 & \quad\! & \times & \times&\times   \\
               \hline
               2 & \quad\! & \times &  \times & \times \\
               \hline
               1 & \quad & \times& \times  &\times   \\
               \hline
               0 & \times & \quad\! & \quad\!  &   \\
               \hline
                 & 0 & 1 & 2 & 3
 \end{array}
\ ,\quad
N_I=11:\;
 \begin{array}{c|c|c|c|c}
               3 & \times & \quad\! & \quad\!&\times   \\
               \hline
               2 & \times & \times &  \quad\! & \times \\
               \hline
               1 & \times & \times& \times  &   \\
               \hline
               0 & \times & \times & \times  &   \\
               \hline
                 & 0 & 1 & 2 & 3
 \end{array}\quad .
$$

\end{document}